\documentclass[aps,prl,superscriptaddress,showpacs,floatfix,notitlepage,twocolumn,preprintnumbers]{revtex4-2}
    \usepackage{amsmath,graphicx,float,csquotes,mdframed,appendix}
\usepackage[colorlinks=true,linkcolor=blue,citecolor=blue, urlcolor=blue]{hyperref}

\newcommand{\trento}{T\raisebox{-0.5ex}{R}ENTo}

\newcommand{\dr}{$\Delta r_{\rm np}$}
\newcommand{\pb}{$^{208}$Pb}
\newcommand{\pbpb}{$^{208}$Pb+$^{208}$Pb}

\begin{document}
 
\title{Determination of the neutron skin of $^{208}$Pb from ultrarelativistic nuclear collisions}


\author{Giuliano Giacalone}
\affiliation{Institut f\"ur Theoretische Physik, Universit\"at Heidelberg,
Philosophenweg 16, 69120 Heidelberg, Germany}

\author{Govert Nijs}
\affiliation{Center for Theoretical Physics, Massachusetts Institute of Technology, Cambridge, MA 02139, USA}

\author{Wilke van der Schee}
\affiliation{Theoretical Physics Department, CERN, CH-1211 Genève 23, Switzerland}
\affiliation{Institute for Theoretical Physics, Utrecht University, 3584 CC Utrecht, The Netherlands}

\begin{abstract}
\noindent Emergent bulk properties of matter governed by the strong nuclear force give rise to physical phenomena across vastly different scales, ranging from the shape of atomic nuclei to the masses and radii of neutron stars. They can be accessed on Earth by measuring the spatial extent of the outer skin made of neutrons that characterises the surface of heavy nuclei. The isotope $^{208}$Pb, owing to its simple structure and neutron excess, has been in this context the target of many dedicated efforts. Here, we determine the neutron skin from measurements of particle distributions and their collective flow in $^{208}$Pb+$^{208}$Pb collisions at ultrarelativistic energy performed at the Large Hadron Collider, which are mediated by interactions of gluons and thus sensitive to the overall size of the colliding $^{208}$Pb ions. By means of state-of-the-art global analysis tools within the hydrodynamic model of heavy-ion collisions,  we infer a neutron skin $\Delta r_{np}=0.217\pm0.058$\,fm, consistent with nuclear theory predictions, and competitive in accuracy with a recent determination from parity-violating asymmetries in polarised electron scattering. We establish thus a new experimental method to systematically measure neutron distributions in the ground state of atomic nuclei.
\end{abstract}

\preprint{CERN-TH-2023-069//MIT-CTP/5558}

\maketitle

Understanding the distribution of neutrons within heavy atomic nuclei has profound implications for our knowledge of the neutron-rich matter that shapes exotic astrophysical objects such as neutron stars. The neutron skin that forms on the surface of heavy nuclei, whereby neutrons are located more diffusely and more on the outside \cite{Trzcinska:2001sy, Zenihiro:2010zz}, represents in particular a sensitive probe of the equation of state (EOS) of neutron matter, whose pressure determines the spatial extent of the neutron distributions. Indeed, nuclear models predict a strong correlation between the neutron skins of heavy nuclei and the masses and radii of neutron stars \cite{Brown:2000pd,Horowitz:2000xj}\@.

While proton distributions in nuclei can be determined in a model-independent way from electron scattering experiments \cite{DeVries:1987atn}, accessing neutron distributions poses a far greater challenge. 
As a consequence, we have only limited experimental constraints on the neutron skin of nuclei, $\Delta r_{np}$, defined as the difference in root mean square (rms) radii between protons and neutrons. 
The doubly-magic nucleus \pb{} ($Z=82$, $N=126$) has both protons and neutrons filling up their respective shells and represents an optimal study subject in this context.
A recent, precise deduction of the neutron skin of \pb{} has been achieved by the PREX collaboration \cite{PREX:2021umo} from the measurement of parity-violating asymmetries in polarised electron scattering. On the side of theory, the first calculation of \pb{} and its neutron skin in the context of \textit{ab initio} nuclear theory was also recently performed \cite{Hu:2021trw}\@.
These results, along with information coming from pulsar and gravitational wave observations, portray a picture of nuclear matter that hints at potential tensions \cite{Fattoyev:2017jql,Reed:2021nqk,Essick:2021kjb}\@.

In this article, we determine the neutron skin of \pb{} from a new type of probe.
We use data collected in \pbpb{} collisions performed at ultrarelativistic energy at the CERN Large Hadron Collider (LHC)\@. These collisions produce short-lived quark-gluon plasma \cite{Braun-Munzinger:2007edi,Teaney:2009qa,Gardim:2019xjs} (QGP), the hot phase of quantum chromodynamics (QCD), which behaves like a near-ideal relativistic fluid \cite{Romatschke:2017ejr,Bernhard:2019bmu} before fragmenting into observable particles. In high-energy scattering, interactions are mediated by gluons, such that the combined distribution of protons and neutrons (altogether called nucleons) within the colliding \pb{} ions determines the shape and the size of the created QGP\@.  
Employing the latest advances in simulation and Bayesian inference tools within the hydrodynamic framework of heavy-ion collisions we reconstruct the geometry of the QGP by using the detected particle distributions. In conjunction with the precise knowledge of the proton density this 
enables us to place a tight constraint on the neutron skin of \pb{}\@.

\begin{figure}[ht!]
    \centering
    \includegraphics[width=\linewidth]{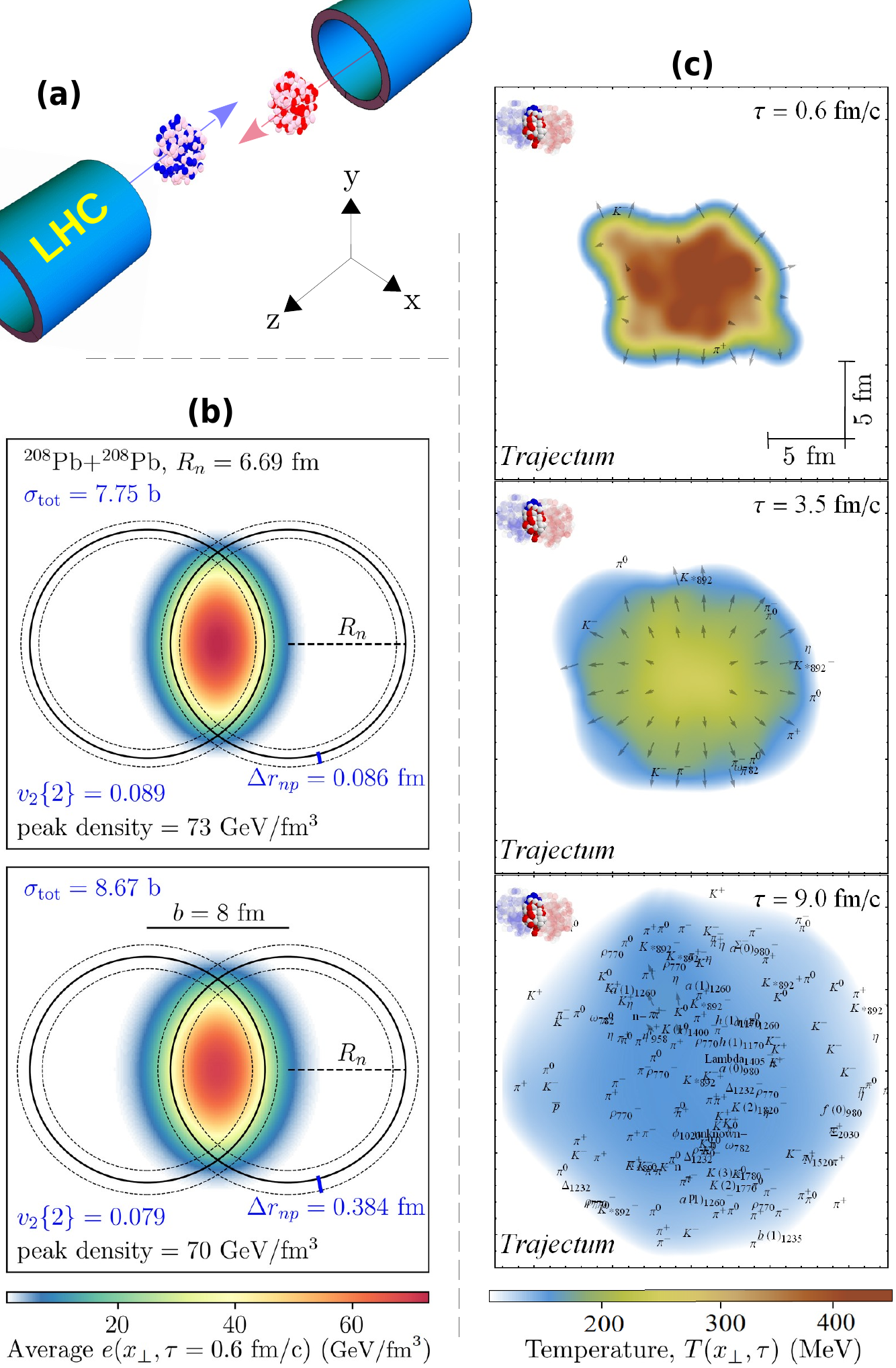}
    \caption{Neutron skin and collective flow in relativistic nuclear collisions. 
    \textbf{a}: Two ions collide with impact parameter $b=8\,$fm. Both ions are Lorentz-contracted by a factor $\gamma\approx2500$, and the relevant dynamics hence effectively takes place in the transverse plane, $x_\perp=(x,\,y)$\@.
    \textbf{b}: The collision deposits energy in the interaction region depending on the extent of the neutron skin of the \pb{} nuclei. We consider $\Delta r_{np}=0.086\,$fm (top) and $\Delta r_{np}=0.384$ (bottom)\@. The neutron skin is varied by keeping the half-width neutron radius, $R_n$, constant while changing the neutron diffuseness, as displayed by the dotted lines (see also Eqn.~\eqref{eq:WS} below)\@. A larger neutron skin leads to a considerably larger total hadronic cross section, $\sigma_{\rm tot}$, and the resulting QGP is in addition more diffuse and less elliptical.
    \textbf{c}: We show a single QGP evolving hydrodynamically and being converted into particles (marked in the figure with their respective symbols) as it cools, while expanding both in $z$ and in the transverse plane. The observation of millions such events leads to characteristic azimuthal anisotropies in the momentum distribution of the produced particles, the most important of which is quantified by the rms value of its second Fourier component, the elliptic flow $v_2\{2\}$, which reflects the ellipticity of the QGP\@.
   }
    \label{fig:1}
\end{figure}

\paragraph{\bf The neutron skin and the quark-gluon plasma -} Our understanding of the QGP formed in \pbpb{} collisions is highly developed thanks to the wealth of experimental data collected in the past decade by all LHC experiments, and in particular by the ALICE experiment dedicated to nuclear physics \cite{ALICE:2022wpnmanual}\@. Following Fig.~\ref{fig:1}, in an ultrarelativistic heavy-ion collision in the lab frame (Fig.~\ref{fig:1}a), interactions of gluons deposit energy density in the area of overlap in the so-called \textit{transverse plane}, perpendicular to the beam direction (Fig.~\ref{fig:1}b)\@. The deposition of energy density depends on the collision's impact parameter $b$, on the structure of the colliding nuclei and on the dynamics of the interaction itself. 
 
Phenomenological studies have established a picture where the colliding ions are treated, in each collision (or \textit{event}), as a superposition of nucleons that participate in the interaction. Both boosted nuclei are thus associated with a profile of matter in the transverse plane, $\mathcal{T_{L,R}}(x_\perp)$, given as the sum of their participant nucleon profiles, typically taken as Gaussians with a width denoted by $w$\@. The interaction process and the subsequent energy depositions are then parameterised following some flexible prescription which can be fine-tuned directly from experimental data. Here we use a T\raisebox{-0.5ex}{R}ENTo-type Ansatz for the energy density of the QGP \cite{Moreland:2014oya, Nijs:2023yab},  
\begin{equation}
\label{eq:e}
e(x_\perp) \propto \left(\frac{\mathcal{T}_L(x_\perp-b/2)^p + \mathcal{T}_R(x_\perp+b/2)^p}{2}\right)^{q/p},    
\end{equation}
where $L,R$ denote the two colliding ions, while $p$ and $q$ are model parameters. As the positions of the participant nucleons shaping the functions $\mathcal{T}_{L,R}$ are sampled in each collision from the neutron and proton densities in the ground state of the scattering ions, the energy density $e(x_\perp)$ is sensitive to their spatial distribution. This can be seen by eye in the density plot of Fig.~\ref{fig:1}b, representing an average energy density over many collisions. The scenario where the colliding \pb{} nuclei have a narrower neutron skin leads to a QGP with a sharper profile over the plane and a higher density peak.

Starting from the initial condition discussed in Fig.~\ref{fig:1}b, the QGP then evolves as a relativistic viscous fluid (with transport properties, such as shear and bulk viscosities, that are also model parameters)\@. For a single event, snapshots of the hydrodynamic expansion obtained using our hydrodynamic code are depicted in Fig.~\ref{fig:1}c. Cooling of the QGP lasts until the confinement crossover is reached, after which at a fixed switching temperature the fluid is converted into a gas of QCD resonance states that can further re-scatter or decay to stable particles. Out of this process, experiments can only detect final event-by-event stable particle spectra, typically denoted by: 
\begin{equation*}
\frac{d^3N_{\rm ch}}{d^2 {\boldsymbol{p}}_T\,d \eta} = \frac{d^2 N_{\rm ch}}{ d p_T\,d \eta} \frac{1}{2\pi} \left (1 + 2 \sum_{n=1}^{\infty} v_n \cos n(\phi-\phi_n) \right ),
\end{equation*}
where $\boldsymbol{p}_T$ is the transverse momentum, $\eta$ is the particle pseudorapidity ($\eta\equiv - \ln \tan ( \theta/2)$ with $\theta$ the polar angle in the $(x_\perp,\,z)$ plane of Fig.~\ref{fig:1}a), and the subscript ch indicates that only charged particles are included. We have conveniently decoupled the spectrum into a distribution of transverse momenta, $p_T\equiv|\boldsymbol{p}_T|$, which quantifies the explosiveness of the QGP expansion, and an azimuthal component developed in Fourier modes, where $v_n$ are the so-called anisotropic flow coefficients that quantify the anisotropy of the particle emission. 
\begin{figure*}[t]
    \includegraphics[width=\linewidth]{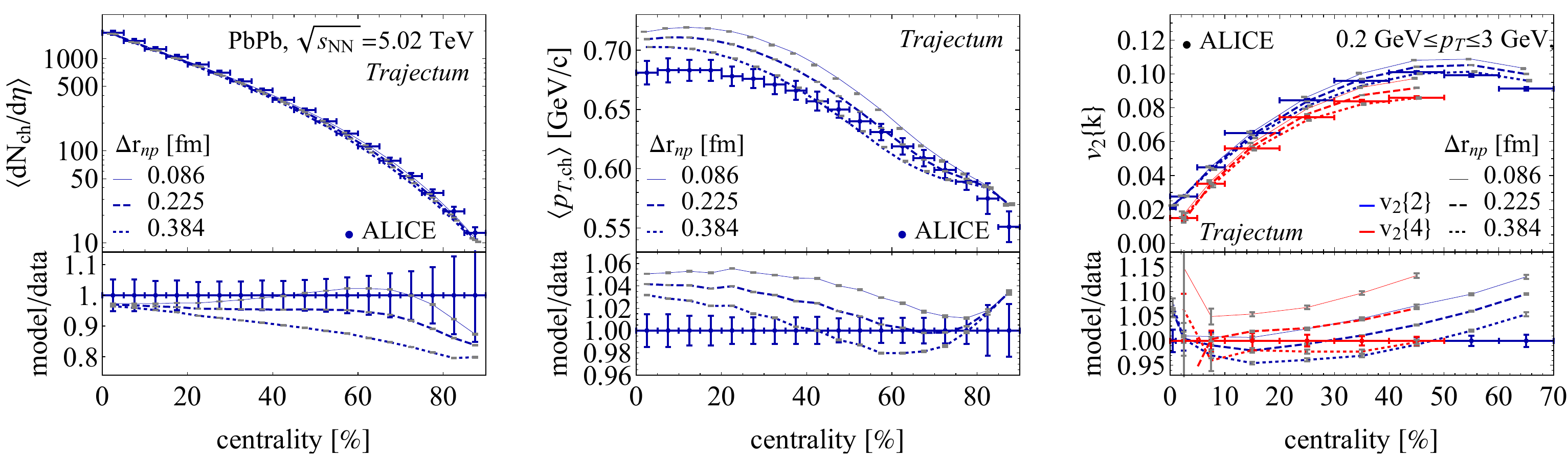}
    \caption{Signature of the neutron skin on bulk particle production in ultrarelativisitic \pbpb{} collisions.
    Varying only the neutron skin size at our optimal parameter settings we show the charged particle multiplicity (left), the mean transverse momentum (middle) and the elliptic flow as measured by $v_2\{k\}$ (right) with a comparison to ALICE data \cite{ALICE:2018ekf, ALICE:2018rtz}\@. A larger neutron skin leads to more collisions, but per collision the multiplicity is lower at larger centralities. The larger size of the QGP leads to a reduced transverse momentum on average. Smearing of the elliptical shape leads to reduced elliptic flow as measured by $v_2\{2\}$ and $v_2\{4\}$\@. Theoretical error bars are statistical only, experimental uncertainties include systematics as well.
    }
    \label{fig:observables}
\end{figure*}

Experimentally the first step is to sort the collisions in centrality classes based on the number of particles that they produce, where 0\% centrality corresponds to events with the highest number of particles at almost zero impact parameter. As a function of centrality one can then measure among others the distributions of $p_T$ and $v_n$ coefficients for different particle species (pions, kaons, protons and more)\@. This generates a wealth of experimental information from which the hydrodynamic model parameters (here, we have 26 in total) can be inferred. The central idea of this manuscript is that of promoting the neutron skin of \pb{} to a model parameter that we constrain from LHC data.

The neutron skin is introduced by considering variations in the neutron diffuseness, $a_n$, in the two-parameter Fermi distributions that model the point-neutron and point-proton densities in the colliding \pb{} nuclei:
\begin{equation}
\label{eq:WS}
\rho_{n,p}(r) \propto \left [ 1 + \exp \left ( \frac{r - R_{n,p}}{a_{n,p}}  \right ) \right ]^{-1}.
\end{equation}
We take $a_p=0.448$\,fm, $R_p=6.680$\,fm (corresponding to an rms proton radius $r_p=5.436$\,fm), and $R_n=6.690$\,fm, which is motivated by the experimental result that the neutron skin is caused by a more diffuse profile rather than a larger half-width radius \cite{Trzcinska:2001sy, Zenihiro:2010zz}\@. 

Before proceeding with a full Bayesian analysis we simulate the QGP formation and evolution for three different values of $\Delta r_{np}$ while keeping all other model parameters fixed.
First, a larger neutron skin leads to a larger total hadronic cross section, $\sigma_{\rm tot}$ (see Fig.~\ref{fig:1}b for an increase from 7.75 to 8.67\,b), because it increases the overall number of events occurring at higher impact parameters.
 
We follow now Fig.~\ref{fig:observables}, showing experimental and model results for quantities that characterise the bulk of particle production from the measured spectra.
The larger $\sigma_{\rm tot}$ for the larger neutron skin induces larger impact parameters at the same centrality. As a consequence, fewer particles are produced for larger values of $\Delta r_{np}$, as clearly visible in the total multiplicity in Fig.~\ref{fig:observables} (left panel)\@. A second effect of a larger skin, highlighted in Fig.~\ref{fig:1}b, is that it leads to more diffuse QGP droplets, which leads to weaker pressure gradients and a slower hydrodynamic expansion. This translates into a lower average momentum of the detected particles, as seen in the middle panel of Fig.~\ref{fig:observables}\@.
In addition Fig.~\ref{fig:1} shows that a larger neutron skin reduces the ellipticity of the QGP\@. This leads to a reduction of the elliptic flow, measured in experiment as a two-particle azimuthal correlation ($v_2\{2\}$, the rms value of the distribution of $v_2$) or as a four-particle correlation ($v_2\{4\}$)\@. Indeed Fig.~\ref{fig:observables} (right) shows the expected reduction and moreover we find that a larger neutron skin enhances the difference between $v_2\{2\}$ and $v_2\{4\}$, which corresponds to larger elliptic flow fluctuations.

\begin{figure}[t]
    \includegraphics[width=\linewidth]{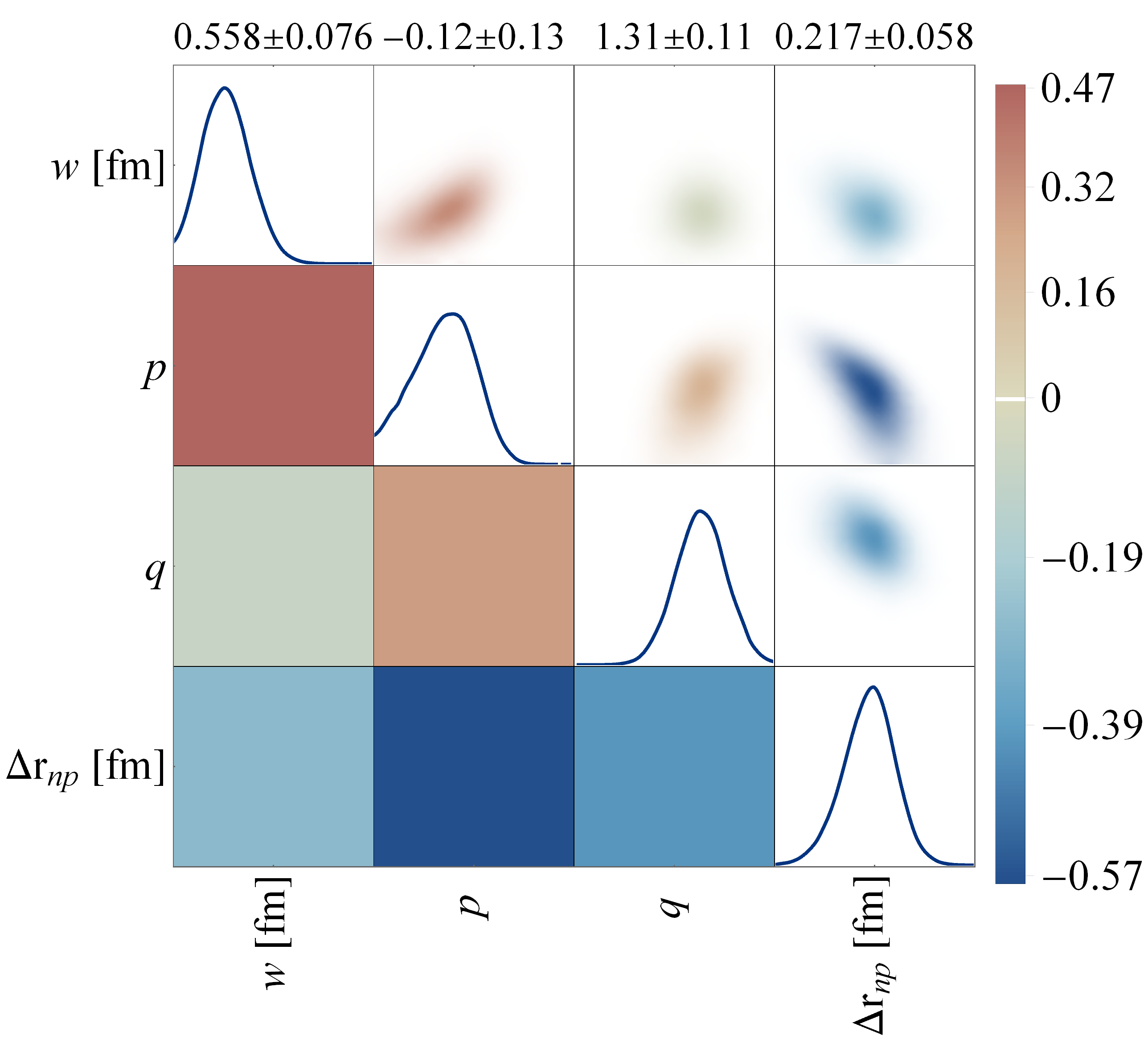}
    \caption{Inferred neutron skin and energy-deposition parameters. We show the posterior distribution of the neutron skin $\Delta r_{np}$, the nucleon width $w$ and the energy deposition parameters $p$ and $q$, together with their expectation values (see top) and correlations. Uncertainties correspond to the standard deviations of the posterior distributions. Especially the $p$ parameter (see Eqn.~\eqref{eq:e}) is highly anti-correlated with $\Delta r_{np}$, as both have a strong effect on the centrality dependence of observables (see also Fig.~\ref{fig:observables})\@.
    }
    \label{fig:3}
\end{figure}

\paragraph{\bf Bayesian inference of the $^{208}$Pb neutron skin -}
Due to the interplay and cross-cor\-re\-la\-tions among parameters and observables, constraining the model from experiment requires advanced Bayesian analysis tools as pioneered in earlier works~\cite{Sangaline:2015isa, Bernhard:2019bmu}\@.
Our analysis makes use of 653 data points in \pbpb{} collisions and a single data point (the total cross section) of proton-nucleus ($p$+\pb{}) collisions. We use 3000 design points for the Gaussian Processes to emulate our collisions as a function of the 26-dimensional parameter space. See the Supplemental Material (SM) 
for a specification of all data, parameters and their inferred distributions.

The posterior distributions are displayed in Fig.~\ref{fig:3} for a subset of parameters that correlate highly with $\Delta r_{np}$\@. These are the parameters appearing in the energy deposition formula, Eqn.~\eqref{eq:e}, namely, the energy deposition parameters $p$ and $q$, as well as the nucleon size, $w$\@. In fact, the $p$ parameter and $\Delta r_{np}$ are the most negatively correlated across our entire parameter space. This is not surprising, as both parameters strongly influence the centrality dependence of observables, whereby a larger neutron skin in particular affects off-central collisions by increasing the total cross section.

Here we briefly revisit Fig.~\ref{fig:observables}, where the middle curve represents our most likely value estimate. In the SM we present the full posterior distributions of our set of 653 data points. There, it can also be seen that the reason for the mismatch between the computed $\langle p_T \rangle$ and experimental data in Fig.~\ref{fig:observables} lies in a slight overestimate of yield of protons, which comes with a larger $p_T$. There is also a significant posterior uncertainty in the anisotropic flow, which is dominated by the emulator uncertainty.

\begin{figure}[t]
    \centering
    \includegraphics[width=0.95\linewidth]{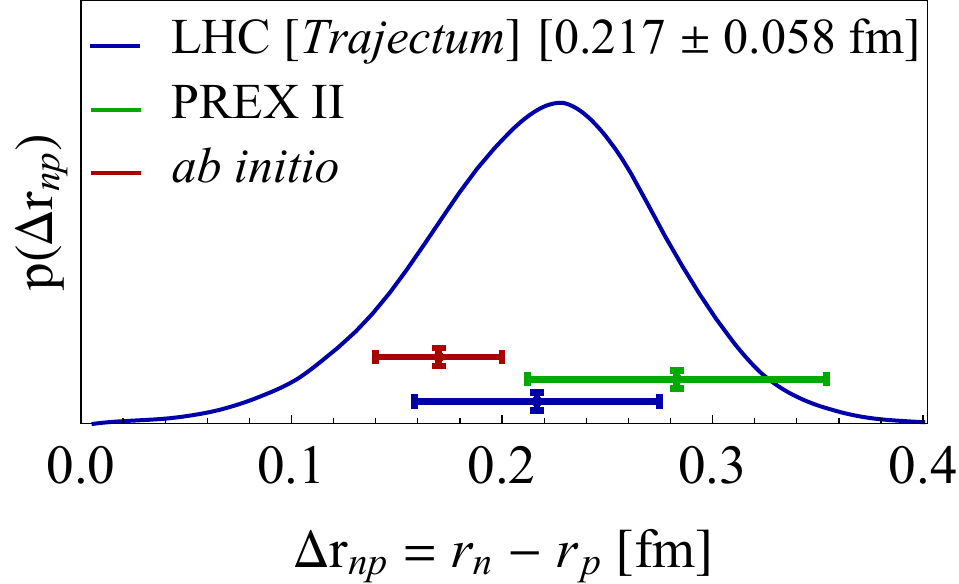}
    \caption{State-of-the-art determinations of the neutron skin of \pb{}. We show the final likelihood distribution of the neutron skin as determined from the LHC data as compared to the values obtained experimentally by the PREX collaboration (including both experimental and theoretical uncertainties in the extraction) \cite{PREX:2021umo} and the estimate of \emph{ab initio} nuclear theory (with an error bar corresponding to a 68\% credibility interval) \cite{Hu:2021trw}\@.
    }
    \label{fig:4}
\end{figure}

In Fig.~\ref{fig:4} we put our new result in context of other state-of-the-art determinations of the skin of \pb{}\@. From the posterior distribution we obtain $\Delta r_{np}=0.217 \pm 0.058\,$fm, corresponding to a point-like rms neutron radius $r_n=5.653 \pm 0.058$ fm. Our result is compatible with both the \textit{ab initio} determination \cite{Hu:2021trw}\@ and the PREX result \cite{PREX:2021umo}, which is competitive in precision. With regards to the EOS of neutron matter, from the relation between $\Delta r_{np}$ and the slope parameter, $L$, of the symmetry energy around the nuclear saturation density \cite{Vinas:2013hua}, we obtain $L=79\pm39\,$MeV, representing the first collider-based constraint on this parameter from high-energy data.

We comment now on the robustness of this result. The total \pbpb{} and $p$+\pb{} cross sections \cite{2204.10148manual,CMS:2015nfb} pose important constraints on the neutron skin. Indeed, excluding these two measurements we obtain $\Delta r_{np} = 0.31 \pm 0.10\,$fm, whereas using exclusively these two data points results in $\Delta r_{np} = 0.03 \pm 0.12\,$fm. Our result comes hence from constraints due to a combination of observables, where the cross section prefers a smaller neutron skin, while other observables prefer a larger value (this is similar for $w$ \cite{Nijs:2022rme})\@. For the first time in Bayesian analyses we include the $\rho_2$ observable \cite{Bozek:2016yoj,ATLAS:2022dovmanual}, a sensitive probe of the initial conditions \cite{Bozek:2020drh,Schenke:2020uqq,Giacalone:2020dln,Giacalone:2021clp,Nijs:2022rme} which measures the correlation between $v_2\{2\}$ and $\langle p_T \rangle$\@. Without this observable, the analysis yields a consistent result, $\Delta r_{np} = 0.243\pm 0.059\,$fm. Also, as introduced in Ref.~\cite{Nijs:2022rme}, we weight the targeted observables according to a prescription that models unknown theoretical uncertainty with respect to $p_T$-differential observables in particular. Turning this weighting off, we find a consistent albeit slightly smaller neutron skin, $\Delta r_{np} = 0.160 \pm 0.057\,$fm.

Further indication of the robustness of our finding comes from the fact that targeting a subset of $p_T$-integrated-only observables, corresponding to 233 ALICE data points, we obtain $\Delta r_{np} = 0.216 \pm 0.057\,$fm. This suggests that the extraction of $\Delta r_{np}$ is likely insensitive to theoretical uncertainties in the particlisation of the QGP at the switching temperature \cite{2011.01430}\@. Lastly, we note that our \trento{} Ansatz of Eqn.~\eqref{eq:e} is very versatile, and may lead to a relatively conservative estimate of the uncertainty on $\Delta r_{np}$\@. Implementing in the future a model of initial conditions motivated by first-principles arguments and with fewer parameters \cite{Heffernan:2023utr}, may lead to stronger constraints than presented here. 

\paragraph {\bf Future skin determinations at the LHC -} 
Albeit computationally expensive, it would be interesting to vary the full neutron profile, by changing $R_n$, or via a multi-parameter function as in Ref.~\cite{Zenihiro:2010zz}. The SM presents an exploratory study of this kind. Increasing the neutron half-width radius does not affect the average multiplicity and the elliptic flow, but leads to a decrease in the mean transverse momentum and a higher hadronic cross section. We speculate, thus, that a more complete analysis could eventually lead to a slightly smaller neutron skin.

We expect our analysis to trigger a program of complementary measurements of skin effects at the LHC\@. A method pioneered by the STAR collaboration utilises the photo-production of vector mesons in ultra-peripheral nucleus-nucleus collisions to infer the average gluon density in the colliding nuclei, and hence the neutron skins \cite{STAR:2022wfe}\@. The extracted skin of $^{197}$Au is in good agreement with nuclear theory predictions \cite{Bally:2023dxi}\@. Therefore,  the same method could be exploited at the LHC to perform an independent extraction of the skin of \pb{}.

In addition, the global analysis presented here uses so-called \emph{soft} observables that depend on particles with transverse momentum of order of the QCD deconfinement temperature, around 150 MeV\@. With high-luminosity LHC runs it may be possible to  constrain the neutron skin as well via \emph{hard} observables, such as high transverse momentum electroweak bosons \cite{Paukkunen:2015bwa, ATLAS:2019ibd}\@. The charge of the produced electroweak bosons can serve as a direct probe of the number of neutron-neutron interactions. By selecting collisions at relatively large impact parameter, it is then possible to determine the dominance of neutrons at the outer edges of the \pb{} nucleus.

It is likely that the nucleus $^{48}$Ca and other ions will be collided at the high-luminosity LHC in the next decade \cite{Citron:2018lsq}\@. This will enable an extended analysis that in particular can be compared with the dedicated CREX measurement of the neutron skin of $^{48}$Ca \cite{CREX:2022kgg}\@. Comparing many different collision systems will furthermore permit us to study ratios of observables that cancel most of the systematic theoretical uncertainties \cite{Xu:2021vpn,Jia:2021oyt,Nijs:2021kvn,Jia:2022qgl}, leading to improved determinations of $\Delta r_{np}$ across the nuclear chart.

\paragraph{{\bf Acknowledgments -}}
We acknowledge discussions with the participants of INT-23-1a, ``Intersection of nuclear structure and high‐energy nuclear collisions'', and the hospitality of the Institute for Nuclear Theory, Seattle. We thank in particular Rituparna Kanungo for interesting discussions. We thank Krishna Rajagopal for comments on the manuscript. G.G.~is supported by the Deutsche Forschungsgemeinschaft (DFG, German Research Foundation) under Germany’s Excellence Strategy EXC 2181/1 - 390900948 (the Heidelberg STRUCTURES Excellence Cluster), SFB 1225 (ISOQUANT) and FL 736/3-1. G.N.~is supported by the U.S.~Department of Energy, Office of Science, Office of Nuclear Physics under grant Contract Number DE-SC0011090.


\bibliographystyle{apsrev4-1}
\bibliography{arxiv, manual}

\newpage

\section*{Supplemental - Bayesian analysis in the \emph{Trajectum} framework}

The \emph{Trajectum} framework consists of an initial stage, a hydrodynamic stage and finally a freeze-out to hadrons that are then evolved by the SMASH code \cite{Weil:2016zrk,dmytro_oliinychenko_2020_3742965,Sjostrand:2007gs}\@. Here we give a brief summary of all parameters involved, how they are constrained, and by which experimental data. All parameters are displayed in boldface. Full details can be found in Ref.~\cite{Nijs:2023yab}\@.

For the initial stage the nucleons are distributed within the colliding nuclei according to Eqn.~2 with parameter $\boldsymbol{a_n}$, whereby each sampled nucleon pair has a minimum distance of $\boldsymbol{d_{\rm min}}$\@. In addition, in Eqn.~2, we consider the half-width radius expanded in spherical harmonics up to the axial quadrupole deformation, $R(\theta)=R(1+ \beta_2 Y_2^0(\theta))$, where $\beta_2$ quantifies the magnitude of the ellipsoidal deformation. Given the structure of \pb{} \cite{Bally:2021qys}, we let $\beta_2$ fluctuate around zero with a standard deviation $\boldsymbol{\sqrt{\langle \beta_2^2\rangle - \langle \beta_2\rangle^2}}$\@. The nucleons that collide are determined by the measured $\sigma_{\rm NN}$ cross section as in \cite{Moreland:2014oya, Nijs:2022rme} and are called participants. Each participant then consists of $\boldsymbol{n_c}$ constituents, each associated with a transverse Gaussian profile with width $v = 0.2\,\text{fm} + \boldsymbol{\chi_\textbf{struct}}(\boldsymbol{w} - 0.2\,\text{fm})$. The center coordinates of the nucleon constituents are distributed according to a Gaussian distribution. This leads to an average Gaussian nucleon profile with a width $\boldsymbol{w}$\@. Superimposition of the nucleon constituent profiles then leads to the thickness functions $\mathcal{T}$ in Eqn.~3\@. The normalisation of each Gaussian profile fluctuates and is equal to $\boldsymbol{N}\gamma/\boldsymbol{n_c}$, where $\gamma$ is sampled from a gamma distribution with mean 1 and width $\boldsymbol{\sigma_\textbf{fluct}}\sqrt{\boldsymbol{n_c}}$\@. The final energy density is then given by Eqn.~3 (main text) with parameters $\boldsymbol{p}$ and $\boldsymbol{q}$\@.

For a time $\boldsymbol{\tau_\textbf{hyd}}$ this energy density then either evolves via free streaming or according to a holographic prescription that starts assuming viscous hydrodynamics (see \cite{Nijs:2023yab})\@. The way this is chosen is with a continuous parameter $\boldsymbol{r_\textbf{hyd}}$ between 0 and 1 which interpolates between free streaming ($\boldsymbol{r_\textbf{hyd}}=0$) and the holographic prescription ($\boldsymbol{r_\textbf{hyd}}=1$)\@.

Next the stress-energy tensor is evolved according to second order hydrodynamics with 9 parameters that determine the transport coefficients, namely $\boldsymbol{\overline{\eta/s}}$, $\boldsymbol{(\eta/s)_\textbf{slope}}$, $\boldsymbol{(\eta/s)_{\delta\textbf{slope}}}$, $\boldsymbol{(\eta/s)_{0.8\,\textbf{GeV}}}$, $\boldsymbol{(\zeta/s)_\textbf{max}}$, $\boldsymbol{(\zeta/s)_{\textbf{m}\times\textbf{w}}}$, $\boldsymbol{(\zeta/s)_{T_0}}$, $\boldsymbol{\frac{\tau_\pi sT}{\eta}}$ and $\boldsymbol{\frac{\tau_{\pi\pi}}{\tau_\pi}}$\@. Of these the first four determine the shear viscosity over the entropy density, $\eta/s$, as a function of the temperature, by fixing the average value, the slope between temperatures of 150 and 300 MeV, the difference in slope at higher temperatures and its value at 800 MeV and higher. The next three determine the maximum of the bulk viscosity over the entropy density, $\zeta/s$, the width of its peak times the maximum, and the location of the maximum, respectively. The last two parameters are dimensionless combinations of first and second order transport coefficients, whereby in particular the first determines the relaxation rate towards first order viscous hydrodynamics. In the hydrodynamic phase the equation of state can vary with the parameter $\boldsymbol{a_\textbf{EOS}}$, which varies $a_n$ in the EOS prescription of \cite{Bernhard:2019bmu} (not to be confused with the parameter $\boldsymbol{a_n}$ discussed in the main text) while keeping the degrees of freedom at high temperatures fixed.

At the switching temperature $\boldsymbol{T_\textbf{switch}}$ the fluid is frozen out to hadrons according to the Cooper-Frye procedure \cite{Cooper:1974mv} with the Pratt-Torrieri-Bernhard (PTB) prescription for viscous corrections \cite{Pratt:2010jt,Bernhard:2019bmu}\@. Lastly, the hadrons are evolved with the SMASH code, whereby we scale all interaction cross sections by a factor $\boldsymbol{f_\textbf{SMASH}}$\@. 

Finally, in the analysis the number of events can vary with a normalising parameter $\boldsymbol{cent_\textbf{norm}}$, which is also the only parameter that has a non-trivial prior likelihood distribution, namely a Gaussian of width 1\%.

\begin{figure*}[t]
    \includegraphics[width=\linewidth]{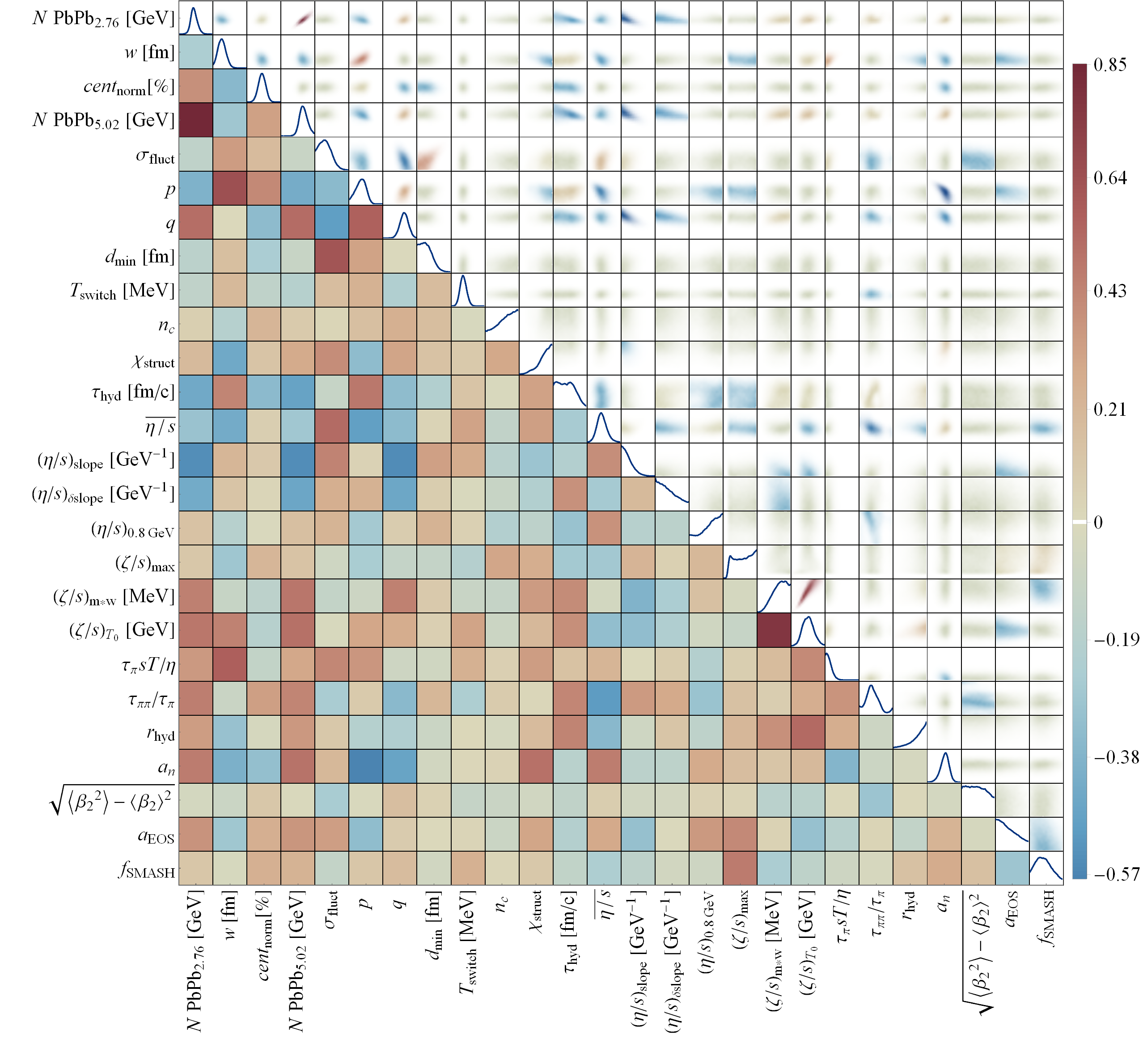}
    \caption{Correlation matrix among all 26 model parameters. Detailed information about the prior ranges is provided in Tab.~\ref{tab:1}.
    }
    \label{fig:correlations}
\end{figure*}

\begin{figure}[t]
    \centering
    \includegraphics[width=0.95\linewidth]{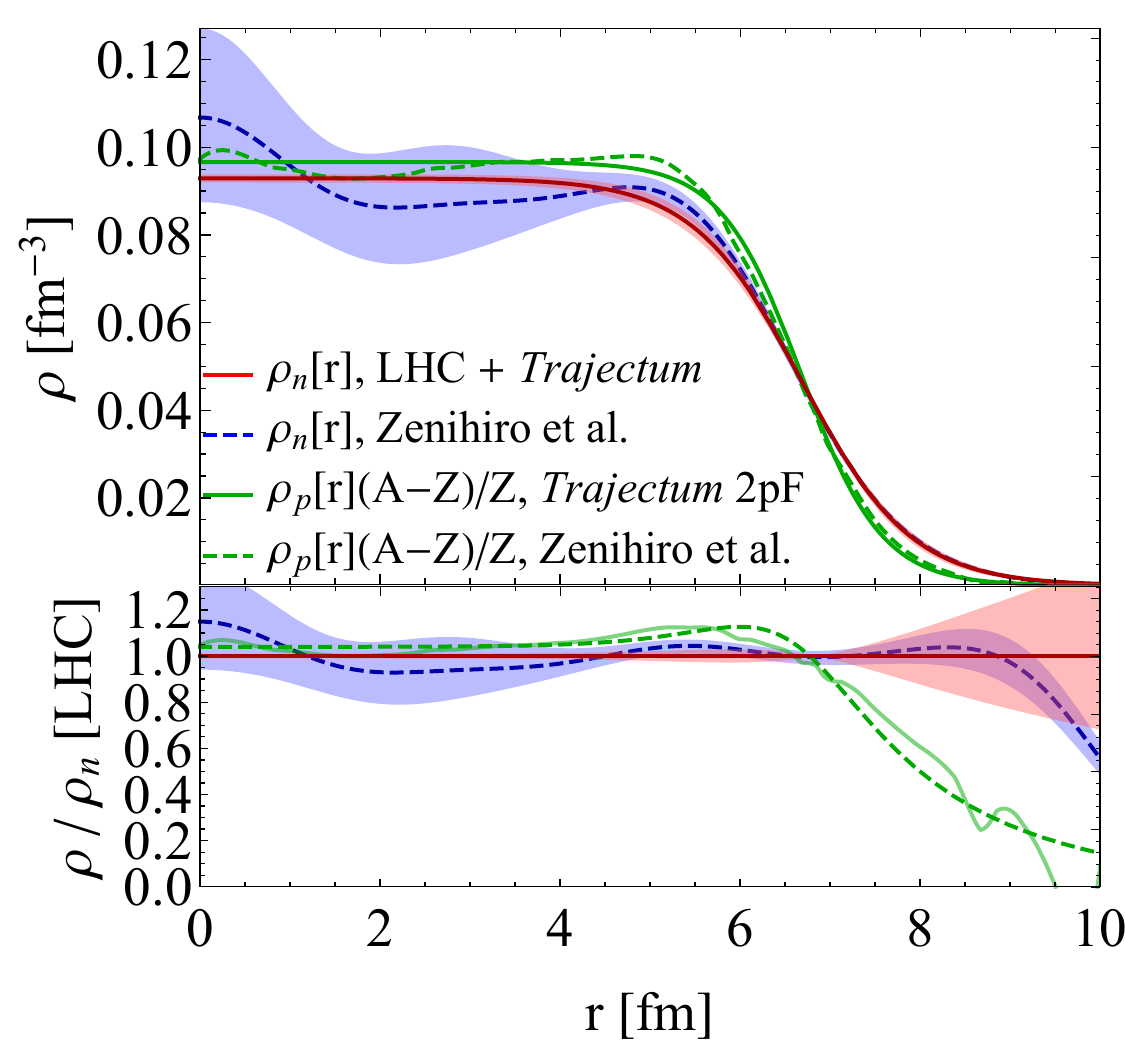}
    \caption{Neutron density measured in polarised proton scattering from Zenihiro et al \cite{Zenihiro:2010zz}\@ compared to that extracted from LHC data. Since we use a single parameter variation ($a_n$ in Eqn.~2) our profile does not vary in the central region. Nevertheless, around the skin at $r \approx 7\,$fm our likelihood band is consistent with \cite{Zenihiro:2010zz}\@. The accurately known proton densities (green) are also in good agreement.
    }
    \label{fig:comparezenihiro}
\end{figure}

We then compare the output of the code with experimental data. In this work we use the hadronic cross section measurements for \pbpb{} and $p$+$^{208}$Pb collisions \cite{2204.10148manual, CMS:2015nfb}, the charged particle yields at 2.76 \cite{ALICE:2010mlf} and $5.02\,\text{TeV}$ \cite{ALICE:2015juo}\@, the identified particle yields $dN_\text{ch}/dy$ and mean transverse momentum $\langle p_T\rangle$ for pions, kaons and protons at $2.76\,\text{TeV}$ \cite{ALICE:2013mez}, as well as unidentified transverse energy $E_T$ \cite{ALICE:2016igk} and fluctuations of mean $p_T$ \cite{ALICE:2014gvd} at $2.76\,\text{TeV}$\@. We also include the integrated anisotropic flow coefficients $v_2\{2\}$, $v_2\{4\}$, $v_3\{2\}$ and $v_4\{2\}$ at both 2.76 and $5.02\,\text{TeV}$ \cite{ALICE:2016ccg} and $p_T$-differential observables with bin boundaries at $(0.25, 0.5, 0.75, 1.0, 1.4, 1.8, 2.2, 3.0)\,\text{GeV}$\@. In particular, this includes spectra for pions, kaons and protons at $2.76\,\text{TeV}$ \cite{ALICE:2013mez}, as well as $v_2\{2\}(p_T)$ for pions, kaons and protons, and $v_3\{2\}(p_T)$ for pions (these data are only available for $p_T > 0.5\,$GeV) \cite{ALICE:2016cti}\@. Finally we also include ``statistically difficult'' observables being the $\rho_2(v_2\{2\}, \langle p_T \rangle)$, $NSC(2, 3)$ and $NSC(2, 4)$ correlators, where $NSC(i, j)$ measures the normalised correlations between $v_i^2$ and $v_j^2$ \cite{ATLAS:2022dovmanual, ATLAS:2019peb}\@.

The posterior for our parameters is then given by Bayes' formula
\begin{equation}
    \mathcal{P}(\boldsymbol{x}|\boldsymbol{y}_{\exp})
    = \frac{e^{-\Delta^2/2}}{\sqrt{(2\pi)^{n} \det\left(\Sigma(\boldsymbol{x})\right)}} \mathcal{P}(\boldsymbol{x}) 
    \label{eq:bayes}
\end{equation}
with $\mathcal{P}(\boldsymbol{x})$ the (flat) prior probability density and where
\begin{equation*}
    \Delta^2
    = \left(\boldsymbol{y}(\boldsymbol{x})-\boldsymbol{y}_{\rm exp}\right)\cdot \Sigma(\boldsymbol{x})^{-1} \cdot \left(\boldsymbol{y}(\boldsymbol{x})-\boldsymbol{y}_{\rm exp}\right),
\end{equation*}
with $\boldsymbol{y}(\boldsymbol{x})$ the predicted data for parameters $\boldsymbol{x}$, $\boldsymbol{y}_{\rm exp}$ the $n$ experimental data points and $\Sigma(\boldsymbol{x})$ is the sum of the experimental and theoretical covariance matrices. The covariance matrices are constructed as in \cite{Bernhard:2019bmu}\@.

The standard procedure is then to run the model at a number of design points in the parameter space as determined by a latin hypercube and use those design points to construct an emulator for $\boldsymbol{y}(\boldsymbol{x})$ to evaluate Eqn.~\eqref{eq:bayes} using the parallel tempered emcee code \cite{B509983H, 1202.3665}\@. In this work we used 3000 design points whereby each design point has 1M initial stages of which we evolve 18k using hydrodynamics and finally simulate about 100k SMASH events to get high statistics even for ultracentral collisions. Every 1 in 15 design points uses 10 times more statistics to allow for ``statistically difficult'' observables that are then emulated using 200 design points. 

The final posterior distributions and their correlations for the standard settings of our fitting procedure are displayed in Fig.~\ref{fig:correlations}\@. We refer to the main text of the paper for details on variations in these settings.

\begin{table}[t]
\centering
 \begin{tabular}{ c c c c  } 
parameter & lower & upper & MAP value \\
 \hline
Norm [2.76 TeV] & 10 & 32 & 17.9 \\
$w$ (fm) & 0.4 & 1 & 0.608 \\
cent$_{\rm norm}$ (\%) & 96 & 104 & 99.6 \\
Norm [5.02 TeV] & 10 & 32  & 22.1 \\
$\sigma_{\rm fluct}$ & 0.1 & 1 & 0.463 \\
$p$ & -0.4 & 0.4 & -0.0526 \\
$q$ & 0.8 & 1.6 & 1.21 \\
$d_{\rm min}$ (fm) & 0 & 1.5 & 0.298 \\
$T_{\rm switch}$ (MeV) & 144 & 162 & 154 \\
$n_{\rm c}$ & 1 & 8 & 2.88 \\
$\chi_{\rm struct}$ & 0 & 1 & 0.932 \\
$\tau_{\rm hyd}$ (fm/$c$) & 0.1 & 0.8 & 0.397 \\
$(\eta/s)_{\rm ave}$ & 0.1 & 0.3 & 0.191 \\
$(\eta/s)_{\rm slope}$ (GeV$^{-1}$) & -1 & 2 & 0.855 \\
$(\eta/s)_{\delta\rm slope}$ (GeV$^{-1}$) & -1 & 1 & -0.393 \\
$(\eta/s)_{\rm 0.8~GeV}$ & 0.08 & 0.4 & 0.344 \\
$(\zeta/s)_{\rm max}$ & 0 & 0.12 & 0.072 \\
$(\zeta/s)_{\rm max\times width}$ & 0 & 0.01 & 0.00865 \\
$(\zeta/s)_{T_0}$ & 0.14 & 0.5 & 0.440 \\
$\tau_\pi T$ & 1 & 10 & 4.06 \\
$\tau_{\pi\pi}/\tau_\pi$ & 0.8 & 6 & 2.52 \\
$r_{\rm hyd}$ & 0 & 1 & 0.798 \\
$a_n$ (fm) & 0.45 & 0.75 & 0.612 \\
$\sqrt{\langle \beta_2^2 \rangle - \langle \beta_2 \rangle^2 }$ & 0 & 0.1 & 0.0217 \\
$a_{\rm EOS}$ & -9.5 & -8.4 & -9.35 \\ 
$f_{\rm SMASH}$ & 0.6 & 1.2 & 0.953 \\
\hline
\end{tabular}
\caption{List of all model parameters constituting the hydrodynamic model used in this paper. A total of 26 parameters are employed. For each parameter, we display the lower and upper boundaries of their prior distributions. The MAP values represent the individual combination of parameters that provides the best fit of the experimental data.}
\label{tab:1}
\end{table}

In Fig.~\ref{fig:comparezenihiro} we compare the resulting posterior likelihood distributions for the proton and neutron densities to measurements, as presented in \cite{Zenihiro:2010zz}\@. Unlike the analysis of \cite{Zenihiro:2010zz}, we vary only $a_n$, so that consequently our neutron density is much more restricted and the proton density is fixed. Nevertheless, in the region where our variation is relevant (i.e., where the diffuseness of the neutron skin matters) the results of the two methods agree remarkably well both in values and in uncertainties.

\begin{figure*}
    \centering
    \includegraphics[width=\linewidth]{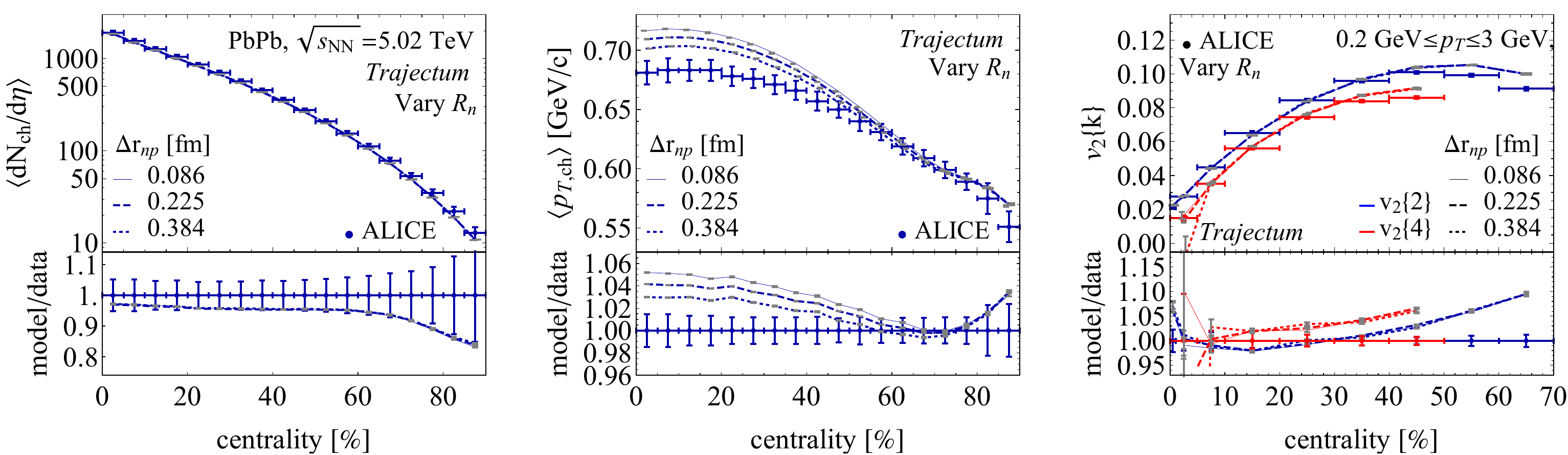}
    \caption{We vary $R_n$ such that the total \dr{} variation matches Fig.~2 in the main text, keeping all other parameters fixed. The size itself has a visible effect on the mean transverse momentum, but little effect on either multiplicity or anisotropic flow.}
    \label{fig:varyingrn}
\end{figure*}

It is an important question what could happen when varying the neutron profile with more parameters, such as done in \cite{Zenihiro:2010zz}. An indication can be found in Fig.~\ref{fig:varyingrn}, where we vary $R_n$ in the Woods-Saxon profile such that the total \dr{} variation matches Fig.~2 in the main text of the paper. It can be seen that the average multiplicity and the elliptic flow remain unaffected, while a smaller $R_n$ results in a larger mean transverse momentum. This is in agreement with the impact of this parameter on the number of participant nucleons and initial energy density eccentricity, which we verified to be also unaffected (note, however, that a smaller nucleus leads to a higher number of binary collisions). It is important that increasing $R_n$ in this figure increases as well the total hadronic cross section from $7.94$ to $8.40\,$b, so that even if the multiplicity and the anisotropic flow remain unaffected, a potential Bayesian analysis would penalize the large $R_n$ by the measurement of $7.67\,$b \cite{Nijs:2022rme,ALICE:2022xir}. Based on this figure a somewhat smaller \dr{} would be possible, but perhaps with a larger bulk viscosity to compensate for the higher mean transverse momentum. A full analysis is however beyond the scope of this work.

\begin{figure*}
    \centering
    \includegraphics[width=.85\linewidth]{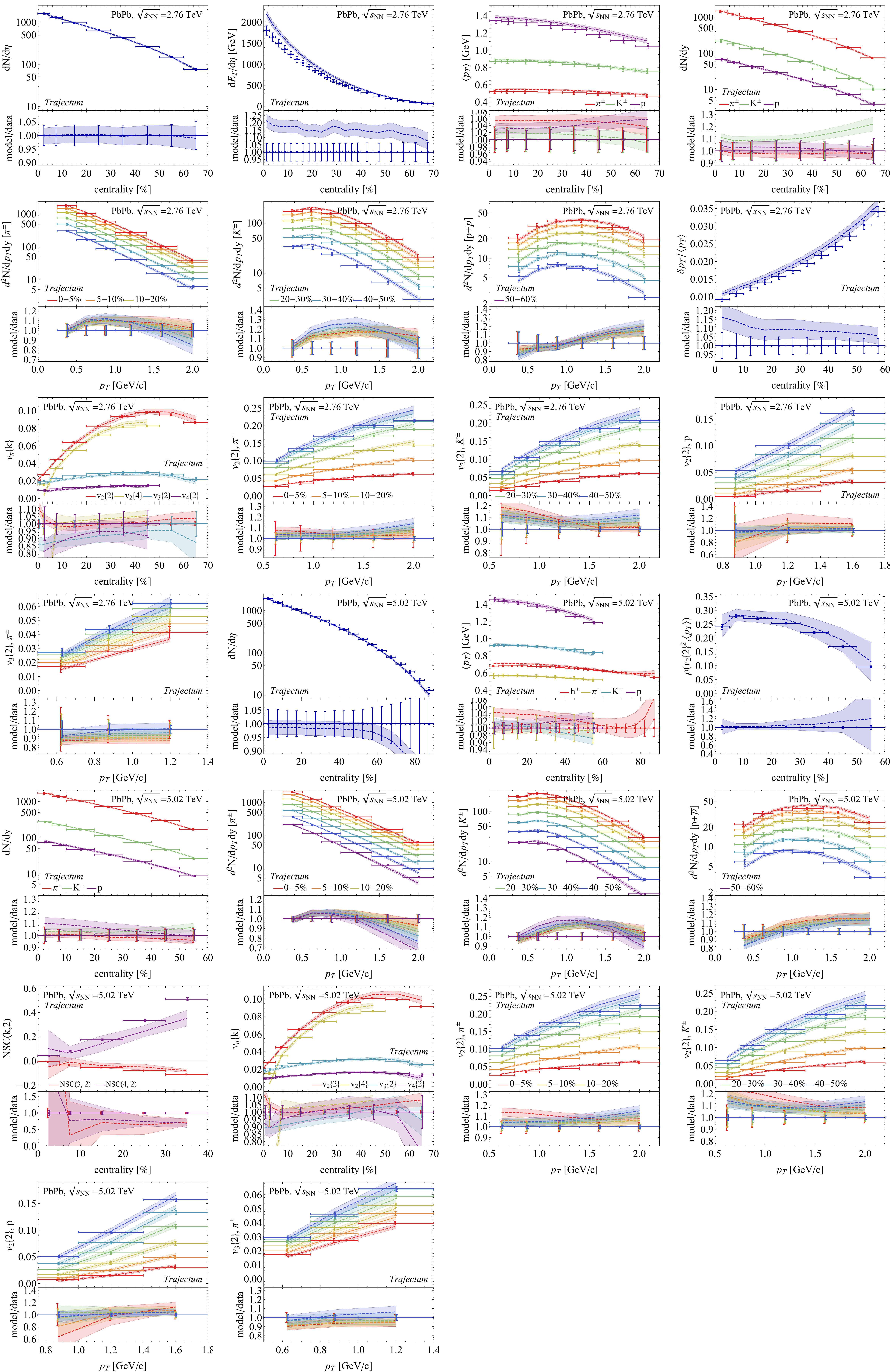}
    \caption{We show all the observables used in our analysis from measurements by the ALICE collaboration \cite{ALICE:2010mlf,ALICE:2015juo,ALICE:2013mez,ALICE:2016igk,ALICE:2014gvd,ALICE:2016ccg,ALICE:2013mez,ALICE:2016cti} and the ATLAS collaboration \cite{ATLAS:2022dovmanual, ATLAS:2019peb}, with the exception of the total hadronic cross sections (see \cite{Nijs:2022rme}). Theoretical uncertainties come from the posterior distribution from Fig.~\ref{fig:correlations}, which includes an emulator uncertainty that is dominant for most flow observables.}
    \label{fig:posteriorobs}
\end{figure*}

Finally, in Fig.~\ref{fig:posteriorobs} we show the posterior results for all the considered observables. Particularly important in relation to the main text is that the mean transverse momentum for identified pions, kaons and protons have an excellent description (row 4, column 3) as opposed to the charged hadron mean transverse momentum in the main text. The reason here is that our model overestimates the number of protons (row 5, column 1) by about 10\%. Given the high proton $p_T$, this small excess leads to the overestimate of experimental data observed in Fig.~2 of the main text. This issue is mainly due to effects in the hadronic phase, such as proton-antiproton annihilation and regeneration (see e.g. \cite{Garcia-Montero:2021haa}), whose detailed inclusion is beyond the scope of \emph{Trajectum}.

\end{document}